\documentclass[sort&compress,review]{elsarticle}

\usepackage[T1]{fontenc}
\usepackage[utf8]{inputenc}
\usepackage{lmodern}
\usepackage{microtype}
\usepackage[english]{babel}
\usepackage{enumitem}
\usepackage{amsmath}
\usepackage{amssymb}
\usepackage{graphicx}
\usepackage{wrapfig}
\usepackage{float}
\usepackage{xcolor}
\usepackage{multirow}
\usepackage{bbm}
\usepackage{gensymb}
\usepackage{textcomp}
\usepackage{gensymb}
\usepackage{lineno}

\usepackage{fancyhdr}
\begin{document}

\begin{frontmatter}
\title{Periodic fluctuations in correlation-based connectivity density time series: application to wind speed-monitoring network in Switzerland}
\author{Mohamed Laib$^1$, Luciano Telesca$^2$ and  Mikhail Kanevski$^1$}
\address{$^1$IDYST, Faculty of Geosciences and Environment, University of Lausanne, Switzerland. \\
$^2$CNR, Istituto di Metodologie per l’Analisi Ambientale, Tito (PZ), Italy \\
Corresponding author: mohamed.laib@unil.ch}

\begin{abstract}

In this paper, we study the periodic fluctuations of connectivity density time series of a wind speed-monitoring network in Switzerland. By using the correlogram-based robust periodogram annual periodic oscillations were found in the correlation-based network. The intensity of such annual periodic oscillations is larger for lower correlation thresholds and smaller for higher. The annual periodicity in the connectivity density seems reasonably consistent with the seasonal meteo-climatic cycle.

\end{abstract}


\end{frontmatter}


\section{Introduction}
\label{intro}
Environmental and geophysical phenomena are characterized by a so complex dynamics that simple statistical tools are not capable of describing the properties of their inner structure that in most cases could be modelled by the interaction of many interconnected dynamical units. Such a vision requires robust methodologies to investigate a phenomenon, in order to capture the dynamical information arising by the interaction of its units. In this context, complex networks represent a powerful tool that allows us to describe the inner structure and the functioning of a wide range of natural as well as technological and social phenomena \cite{da_Fontoura_Costa2011}, where the intensity of such inner interactions plays a dominant role in the underlying dynamics.

Network analysis is an interdisciplinary field, which consists of statistical and computing methodology to study relationships between units (nodes) which have simultaneous behaviour. It was introduced first in sociology and psychology by Moreno \cite{moreno}.
In recent years, complex network analysis have been developed deeply \cite{Barabasi2016, shlomo2010}. Its application has been extended to various scientific fields, such as, Internet and world wide web in computer science, food webs, gene expression, biological neural networks, citation networks in social science \cite{Watts1998, newman2003, barabasi2002}; and it has contributed to gain insight into the nature of complex systems, thus leading to a better understanding of dynamical processes in systems whose elements are not trivially connected \cite{BOCCALETTI2006}.

In environmental sciences, complex networks were especially applied to climatic systems \cite{Tsonis2006, Tsonis2008, Donges2009a, Donges2009b, Gozolchiani2008, Tsonis2004, Yamasaki2008}. The use of climate networks has shed light on several important features characterizing climate systems \cite{Tsonis2006, Tsonis2008, Swanson2009, Donges2009a, Donges2009b, Gozolchiani2008, Tsonis2004, Yamasaki2008, Runge2014}. As an example, atmospheric teleconnections, whose dynamics are not well understood yet, were investigated by using climate networks; it was found, in particular, that teleconnections in the extra-tropics play the role of super-nodes in the corresponding networks, making climate more stable and more efficient in transferring information \cite{Tsonis2008}. The network-based investigation of teleconnections disclosed also some features of extreme climate events, such as the El Niño-Southern Oscillation (ENSO) \cite{Tsonis2008, Gozolchiani2008}. The most common approach in constructing a climate network is based on the gridding of a given climate field, where each grid cell is a node, and edges between nodes describe statistically significant relationships using some linear or nonlinear correlation metrics \cite{Tsonis2004, Donges2009b}. In this approach, edges characterized by statistical significance lower than a certain threshold are simply disregarded and all the remaining edges are equally weighted, leading to an unweighted network \cite{Tsonis2008, Donges2009b, Steinhaeuser2009}. 

The multitude of studies on wind speed was focused on the analysis of wind speed as single time series, to which different statistical techniques were applied, such as distributional analysis  \cite{Shipkovs2013}, data mining \cite{Astolfi2016}, non-linear data driven models based on machine learning algorithms \cite{DeGiorgi2014, Douak2013}, fractal analysis \cite{DEOLIVEIRASANTOS2012, Fortuna2014}, multifractal analysis \cite{Telesca2016}. Up to our knowledge, no network-based analysis has been performed on wind speed field so far.

Our study aims at proposing, for a wide wind speed-monitoring network in Switzerland, a time-varying network analysis, in which each monitoring wind speed station represents a node, and the edges among nodes represent 
the Pearson correlation coefficient among the wind speed time series recorded within a certain time interval $\delta T$; shifting $\delta T$ through time, the temporal variation of properties of the network can be, then, investigated. 

Similarly to the above mentioned network approaches, after fixing a threshold, we analyse the collective behaviour of all those edges that are featured by a correlation coefficient, which is lower or higher than the threshold. The performed analysis, by varying also the threshold, would allow us to depict more deeply the collective behaviour of the wind speed network, not only in the time domain but also in the "correlation domain".

\section{Data and exploratory analysis}
\label{sec:1}
\begin{figure}[h]
\centering
\includegraphics[width=\linewidth]{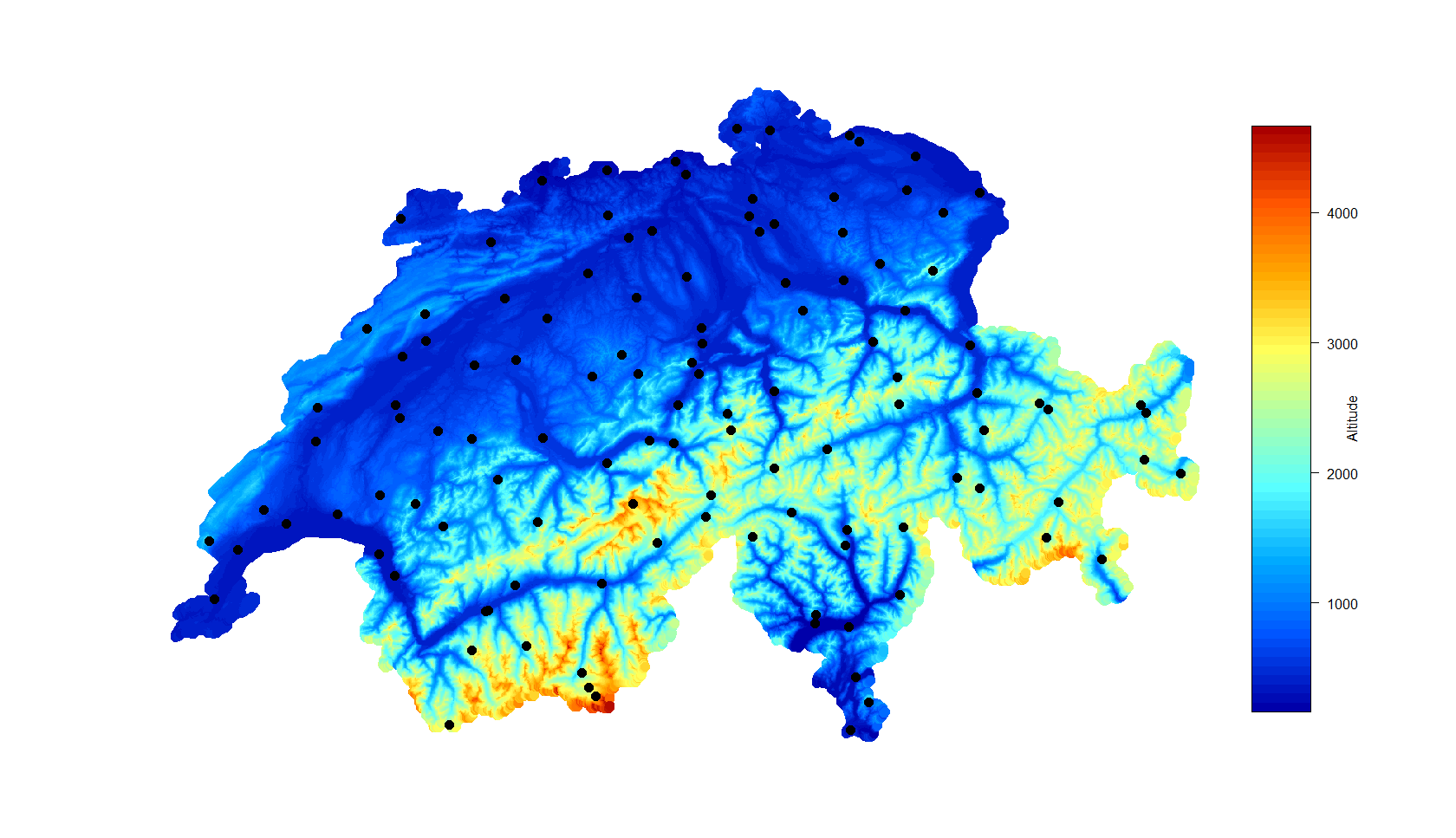}
\caption{ \footnotesize Study area and location of the $119$ used wind measurement stations.}
\label{fig:1}  
\end{figure}

\begin{figure}
\centering
\includegraphics[width=\linewidth]{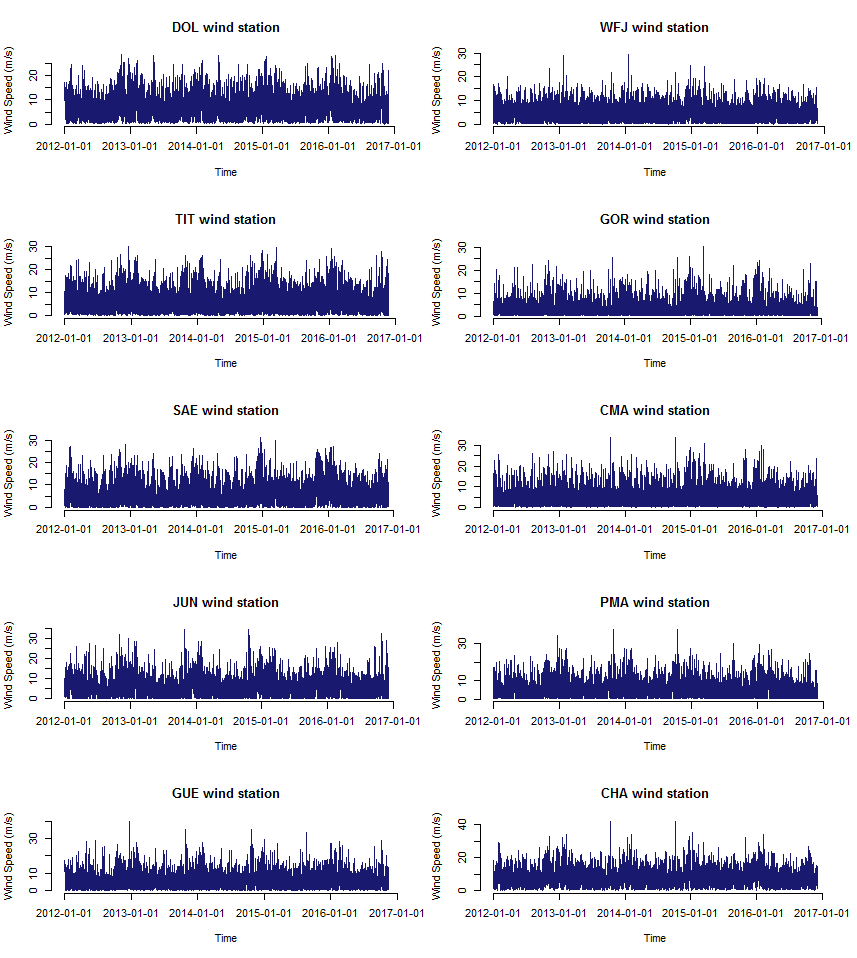}
\caption{ \footnotesize Time series of wind speed from some measuring stations.}
\label{fig:2}  
\end{figure}

In Switzerland, a very dense network of high frequency wind speed stations is operating. The data, which consist of long-term wind speed measurements recorded with a sampling time of 10-minutes, are stored by the Federal Office of Meteorology and Climatology of Switzerland (IDAWEB, MeteoSwiss). Among more than $400$ measuring stations, in this study we selected only $119$ for their low percentage of missing data (the total percentage is $8\%$ data, and the maximum percentage of missing data per day is about $10\%$), during the observation period from $2012$ to $2016$. Fig. \ref{fig:1} shows the location of the $119$ measurement stations used in this study. Fig. \ref{fig:2} shows, as an example, some wind speed time series. We firstly performed a distributional analysis to identify the distribution that better describes our wind speed data. The distributions that are generally employed to fit wind speed data are Weibull \cite{use1}, Gamma \cite{use2}, and Generalized Extreme Values (GEV) \cite{GEV1}. 
\begin{itemize}
\item The Weibull distribution is a two parameters distribution defined as  \cite{reisx}:

\begin{equation}
f(x;\lambda,k)=\left\{\begin{array}{rcl}
\frac{k}{\lambda} (\frac{x}{\lambda})^{k-1} e^{-(\frac{x}{\lambda})^{k}} \qquad x\geq 0\\
\\
0   \qquad \qquad  \qquad x < 0
\end{array}\right.  
\label{wb}
\end{equation}
where $k$ is the shape parameter and $\lambda$ is the scale.

\item The Gamma distribution is also a two parameters distribution and is defined as:
\begin{equation}
f(x;\alpha,\beta)=\frac{\beta^{\alpha}x^{\alpha-1}e^{-x\beta}}{\Gamma(\alpha)} \qquad for \; x \geq0 \; and \; \alpha,\beta > 0
\end{equation}
where $\alpha$ is the shape and $\beta$ is the rate of the Gamma distribution \cite{qpEVT}.

\item The GEV distribution combines the three subfamilies of Gumbel, Fréchet and Weibull distribution and is defined as \cite{colesB}:
\begin{equation}
F(x;\mu, \sigma, \xi ) = exp\{ -[1+\xi(\frac{x-\mu}{\sigma})]^{\frac{-1}{\xi}} \} 
\end{equation}
where $ 1+ \xi(x-\mu)/\sigma > 0$, and $\mu $ is the location parameter, $\sigma $ is the scale parameter and $\xi $ is the shape.
\end{itemize}

In order to evaluate the goodness-of-fit of the data with each probability distributions, we used the well-known Kullback-Leibler divergence (KL). Given a random sample $X_1, \ldots, X_n$ from a probability distribution $P(x)$ with density function $p(x)$ over a non-negative support, if we suppose that the sample comes from a specific probability distribution $Q(x)$ with a density function $q(x)$, the KL information on the divergence between $P(x)$ and $Q(x)$  is given by the following formula \cite{kullback1951}: 

\begin{equation}
D_{KL}(p\|q)=\int_0^\infty p(x) ln \frac{p(x)}{q(x)}dx.
\end{equation}

The entropy of the distribution $P(x)$ is defined as
\begin{equation}
H(P)=-\int_0^\infty p(x) \, ln \, p(x) \, dx
\end{equation}
which can be estimated from the sample \cite{vasicek1976}. 
\begin{equation}
\widehat{H}_{mn}=\frac{1}{n}\sum_{i=1}^{n} \, ln \{\frac{n}{2m}[X_{(i+m)}-X_{(i-m)}]\}
\end{equation}
where $m$ is the window size smaller than $\frac{n}{2}$. 

As an example, if we consider that $Q(x)$ is a Gamma $(\alpha, 1)$ distribution:
\begin{equation}
\int_0^\infty p(x) \, ln \, q(x) \, dx = - ln \, \Gamma(\alpha)-E(X)+(\alpha - 1) E(ln X)
\end{equation}
then the KL information can be given by:
\begin{equation}
D_{KL}(p\|q)=- \widehat{H}_{mn} + ln\Gamma(\alpha) + \bar{X}-(\alpha -1)\overline{lnX}
\end{equation}
where $\bar{X}=\frac{1}{n}\sum X_i$ and $\overline{ln X}= \frac{1}{n}\sum ln X_i$ \cite{waal1996}.

It is known that the information divergence $D_{KL}(p\|q) \geqslant 0$. Therefore, if $D_{KL}(p\|q)=0$, the sample comes from the specific probability distribution $Q(x)$ \cite{Ebrahimi1992, waal1996}.

\begin{figure}
\centering
\includegraphics[width=\linewidth]{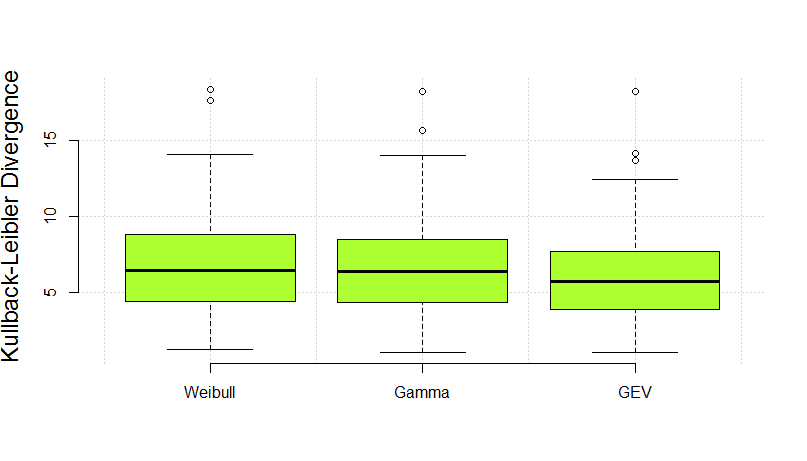}
\caption{ \footnotesize Values of the Kullback-Leibler Divergence.}
\label{fig:4}  
\end{figure}

In each case and for each time series we calculated the KLD and, after averaging all the KLD-values, we obtained the results shown in Fig. \ref{fig:4}: it is suggested, then, that the GEV distribution performs slightly better than the other two distributions.

\section{Correlation networks}
The correlation-based network of the wind speed data can be presented as graph, in which each measuring station is a node \cite {cit_2}; the edges of the network are weighted by the Pearson correlation coefficient \cite{correlation}, defined as:
\begin{equation}
\rho_{XY}=\frac{\sum(x_{i}-\bar{x})\: \sum(y_{i}-\bar{y})}{\sqrt{{\sum(x_{i}-\bar{x})}^2} \: \sqrt{{\sum(y_{i}-\bar{y})}^2}}
\end{equation}
where $\bar{x}$, $\bar{y}$ denote the mean of $X$, $Y$ respectively , which represent $2$ wind speed time series. 

The coefficient of correlation, as it is well known, represents the simplest statistical measure used to evaluate the interdependence between two random variables.

Fixing a threshold for the correlation coefficient, only a subset of the edges of the network will be taken into account. Changing the threshold, of course, will change the network's topology that depends on the correlation. Since we are dealing with time series, even the network's topology will change through time; in fact, dividing the entire observation period into windows of a certain duration, we can construct in a certain window a network, whose topology, given by the multitude of interconnected nodes, will be different from that constructed in another different window. Among the several quantities that can be defined and used for networks, in this study we focus on the connectivity density, defined as follows:

\begin{equation}
\Delta= \frac{E}{N(N-1)/2}
\label{density}
\end{equation}
where $E$ is the number of edges whose correlation coefficient has a certain relationship with a threshold and $N$ is the number of nodes. If $\Delta=0$, all nodes are not connected; if $\Delta=1$, all nodes are connected. According to this measure, each network can be identified by a value of $\Delta$ between $0$ and $1$.

\begin{figure}
\centering
\includegraphics[width=\linewidth]{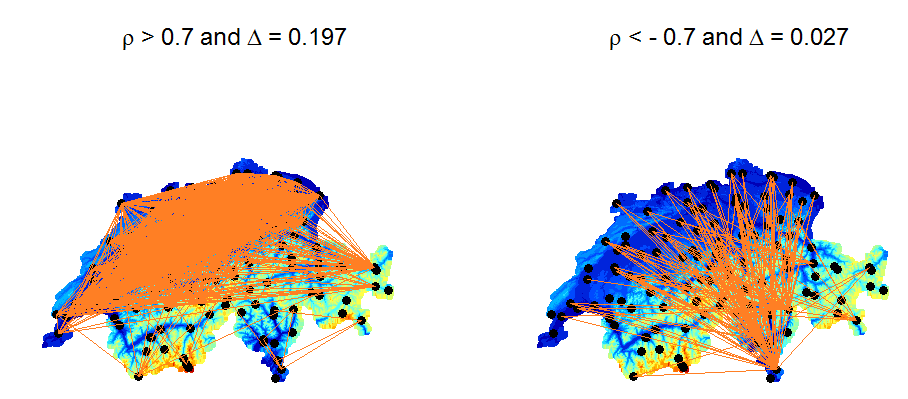}
\caption{ \footnotesize Network of wind speed stations for one day (01-04-2014). Left: positive correlation of 0.7 threshold. Right: negative correlation of -0.7 threshold.}
\label{fig:6}  
\end{figure}

The procedure of constructing the network is described below:
\begin{enumerate}
\item All the wind time series are divided into windows of duration of 1 day each;
\item In each window, we construct a network for a certain threshold (see Fig. \ref{fig:6});
\item The connectivity density is then computed by using formula in  Eq \ref{density};
\item A daily time series of connectivity density is formed.
\end{enumerate}

One issue could be given by the use of the Pearson correlation coefficient instead of any other non-linear correlation measure, since the interactions among the wind speed series could be non-linear. Donges et al. \cite{Donges2009b} examined this aspect and found that the networks constructed by means of the Pearson correlation coefficient and that by mutual information are significantly equivalent; thus, in our case, it is much effective to use the simplest possible correlation measure that is the Pearson correlation.

A crucial point in the construction of the network is the selection rule for threshold on the correlation coefficient. In previous studies, it was fixed a rather high threshold for the absolute value of the correlation coefficient, and all the edges whose absolute correlation coefficient was above the threshold were kept for the construction of the network; in this manner, all the edges weighted by positive or negative correlation coefficient were considered as equally contributing \cite{Donges2009a, Peron2014, Steinhaeuser2011, Steinhaeuser2012}. However, positive correlation indicates that the two nodes evolve in phase, while negative correlation indicates that they evolve in opposition of phase, and this suggests that actually the interaction between the involved nodes is not the same.

\begin{figure}
\includegraphics[width=\linewidth]{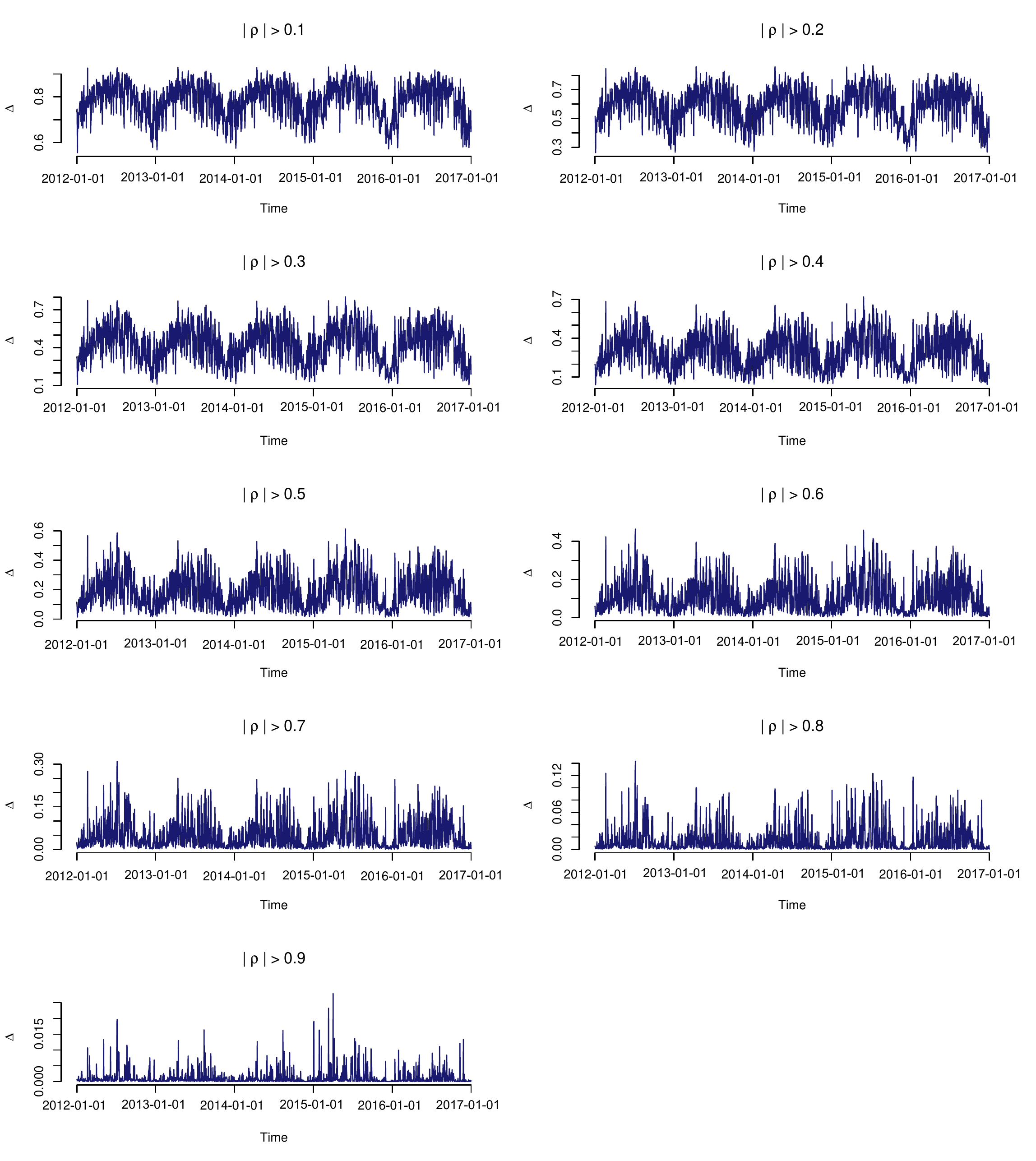}
\caption{ \footnotesize Time series of daily  connectivity density; for different thresholds of the absolute value of the correlation.}
\label{fig:8b}  
\end{figure}

Fig. \ref{fig:8b}  shows the daily time series of the connectivity density concerning our wind monitoring network for different thresholds of the absolute value of the correlation. As it is clearly found by visual inspection, the daily time series of the connectivity density is characterized by an annual periodicity for low correlation thresholds, while it appears more intermittent for higher thresholds,  although a weak yearly amplitude modulation can be still observed.
A more detailed investigation of the network's topology can be performed considering the weights with their sign, simply because positive and negative correlation relies with different types of interaction among the nodes. 

Fig. \ref{fig:7} and Fig. \ref{fig:8} show the daily time series of the connectivity density for different positive and negative correlation thresholds. Similarly to Fig. \ref{fig:8b}, also for these time series the annual periodicity is enhanced for low correlation thresholds, but becomes less intense with the increase of the threshold.

Since the goal of this study is to analyse the periodic behaviour of the time series of connectivity density for the wind speed-monitoring network in Switzerland, we adopted a very robust method for the identification of cycles that was recently proposed by Ahdesmäki et al. \citep{Ahdesmaki2005} the correlogram-based robust periodogram, described in the following section.

\section{The Robust periodogram}
Considering the simplest model of a periodic time series, like the following
\begin{equation}
y_n= \beta cos(\omega t + \phi) +\varepsilon_n
\label{eqp1}
\end{equation}
where $\beta>0$  is a constant, $0<\omega<\pi$, $\phi$ is a uniform random variable in ($-\pi$, $\pi$], and $\{\varepsilon_n\}$ is a sequence of uncorrelated zero-mean random variables with variance $\sigma^2$ that does not depend on $\phi$, the well-known formula of the classical Fourier-based periodogram is given by:
\begin{equation}
I(\omega)= \frac{1}{N} \bigg| \sum_{n=1}^{N} y_{n}e^{-\omega n}\bigg|^2, \quad 0\leq \omega \leq \pi
\label{eqp2}
\end{equation}
where $N$ is the length of the time series. The periodogram is evaluated at normalized frequencies

\begin{equation}
\omega_1= \frac{2\pi l}{N}, l=0,1,\ldots,a
\label{eqp3}
\end{equation}
where $a=[(N-1)/2]$ and $[x]$ indicates the integer part of $(N-1)/2$ and $x$ respectively. Assuming that  the amplitude of the series is modulated significantly by sine of frequency $\omega_0$, then the periodogram  is very probably peaked at that frequency. Otherwise, if the time series is a realization of uncorrelated process with $\beta=0$ in Eq. \ref{eqp1}, then the periodogram appears uniform and flat at any frequency bands \cite{Priestley1981}. Ahdesmäki et al. \citep{Ahdesmaki2005} proposed a robust detection of periodicities on the basis of the estimation of the autocorrelation function. The periodogram is equivalent to the correlogram spectral estimator

\begin{equation}
S(\omega) = \sum_{k=-N+1}^{N-1}\widehat{r}(k)e^{-i \omega k}
\label{eqp4}
\end{equation}
where

\begin{equation}
\widehat{r}(m)=\frac{1}{N} \sum_{k=1}^{N-m}y_{k} y_{k+m}
\label{eqp5}
\end{equation}
is the biased estimator of the autocorrelation function The sample correlation function between two sequences with length $N$ is given by:

\begin{equation}
\rho(m)=\frac{\frac{1}{N} \sum_{1=1}^{N} (x_{i}-\bar{x})(y_{i}-\bar{y})}{\sigma_{x} \sigma_{y}}
\label{eqp6}
\end{equation}
where the bar over the symbol indicates the mean. On the base of the relationship between the estimator of the autocorrelation function $\widehat{r}(m)$  and the estimator of correlation function $\widetilde{\rho}(m)$  between the sequences $y_k$  and $y_{k+m}$, Ahdesmäki et al. obtained the following robust spectral estimator

\begin{equation}
\widetilde{S}(\omega)=2\Re\left( \sum_{k=0}^{L}\widetilde{\rho}(k)e^{i \omega k}
\right) - \widetilde{\rho}(0)
\label{eqp7}
\end{equation}
where $\Re(x)$ is the real part of $x$. $L$ is the maximum lag for which the correlation coefficient $\widetilde{\rho}$ is computed. It was shown that the performance of this spectral estimator is better than that of the standard periodogram in identifying the periodicities in a wide range of types of time series, short, with outliers, with added noise, and linear trends \citep{Ahdesmaki2005}.

\section{Results and discussion}

\begin{figure}
\includegraphics[width=\linewidth]{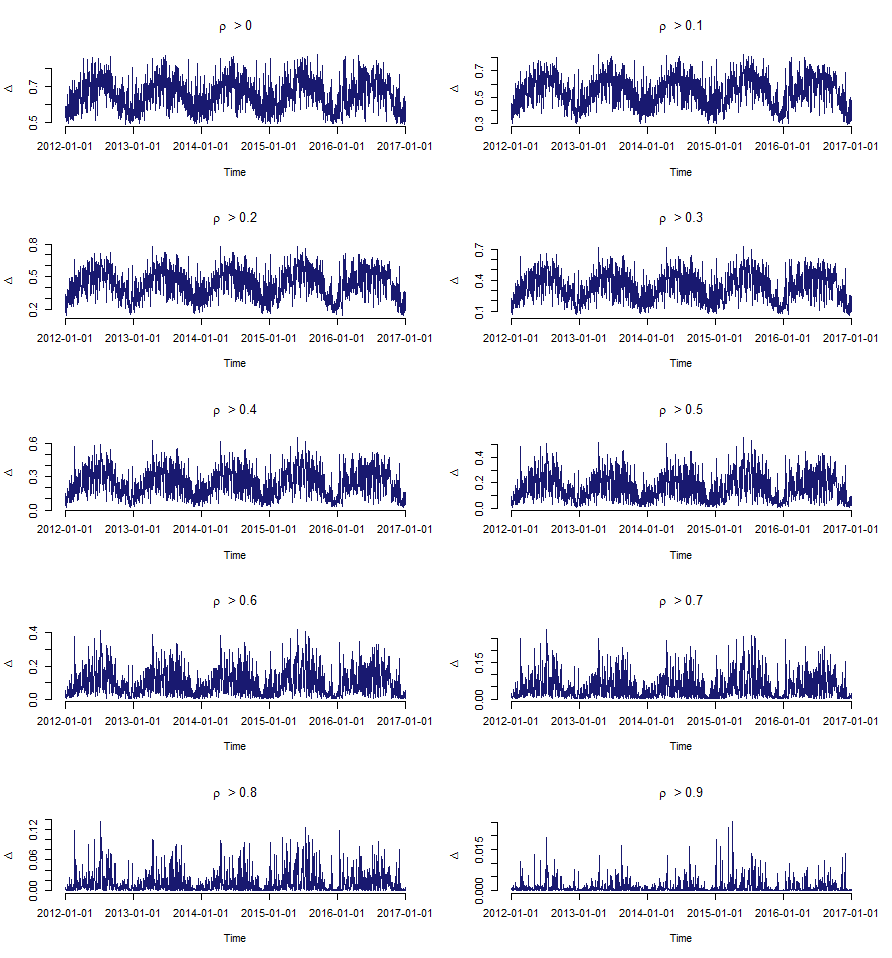}
\caption{ \footnotesize Time series of daily connectivity density (positive correlation thresholds).}
\label{fig:7}  
\end{figure}

\begin{figure}
\includegraphics[width=\linewidth]{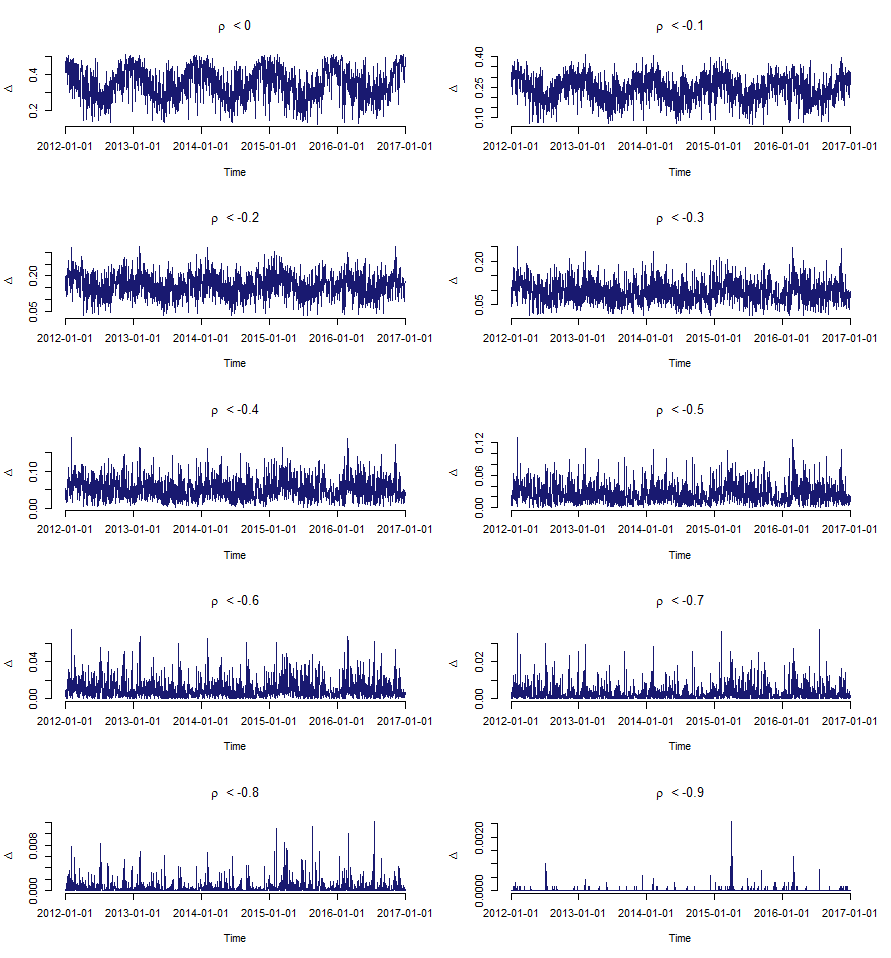}
\caption{ \footnotesize Time series of daily connectivity density (negative correlation thresholds).}
\label{fig:8}  
\end{figure}

\begin{figure}
\includegraphics[width=\linewidth]{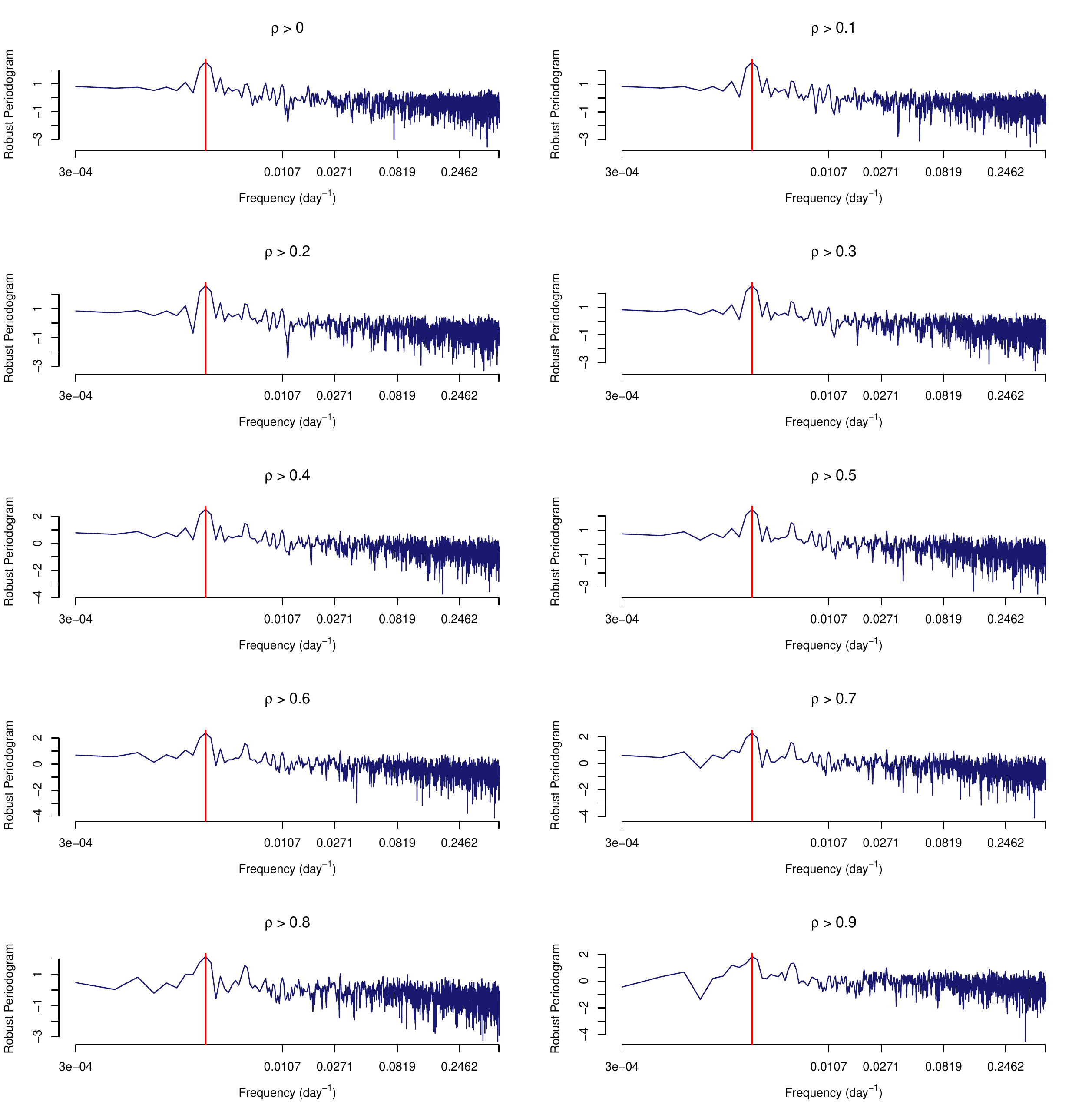}
\caption{ \footnotesize Robust periodogram for positive correlation.}
\label{fig:9}  
\end{figure}

\begin{figure}
\includegraphics[width=\linewidth]{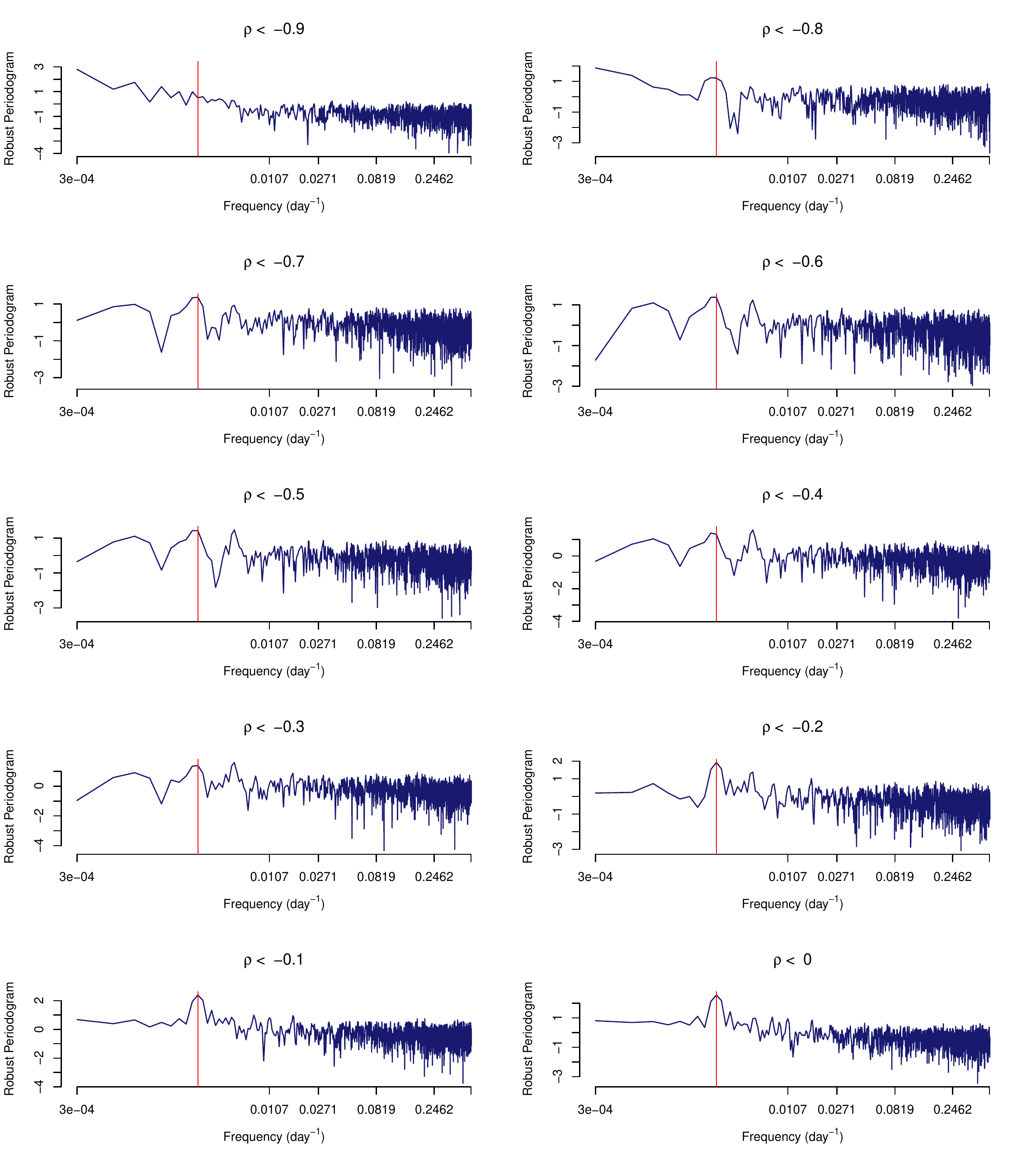}
\caption{\footnotesize Robust periodogram for negative correlation.}
\label{fig:10}  
\end{figure}

\begin{figure}
\centering
\includegraphics[width=\linewidth]{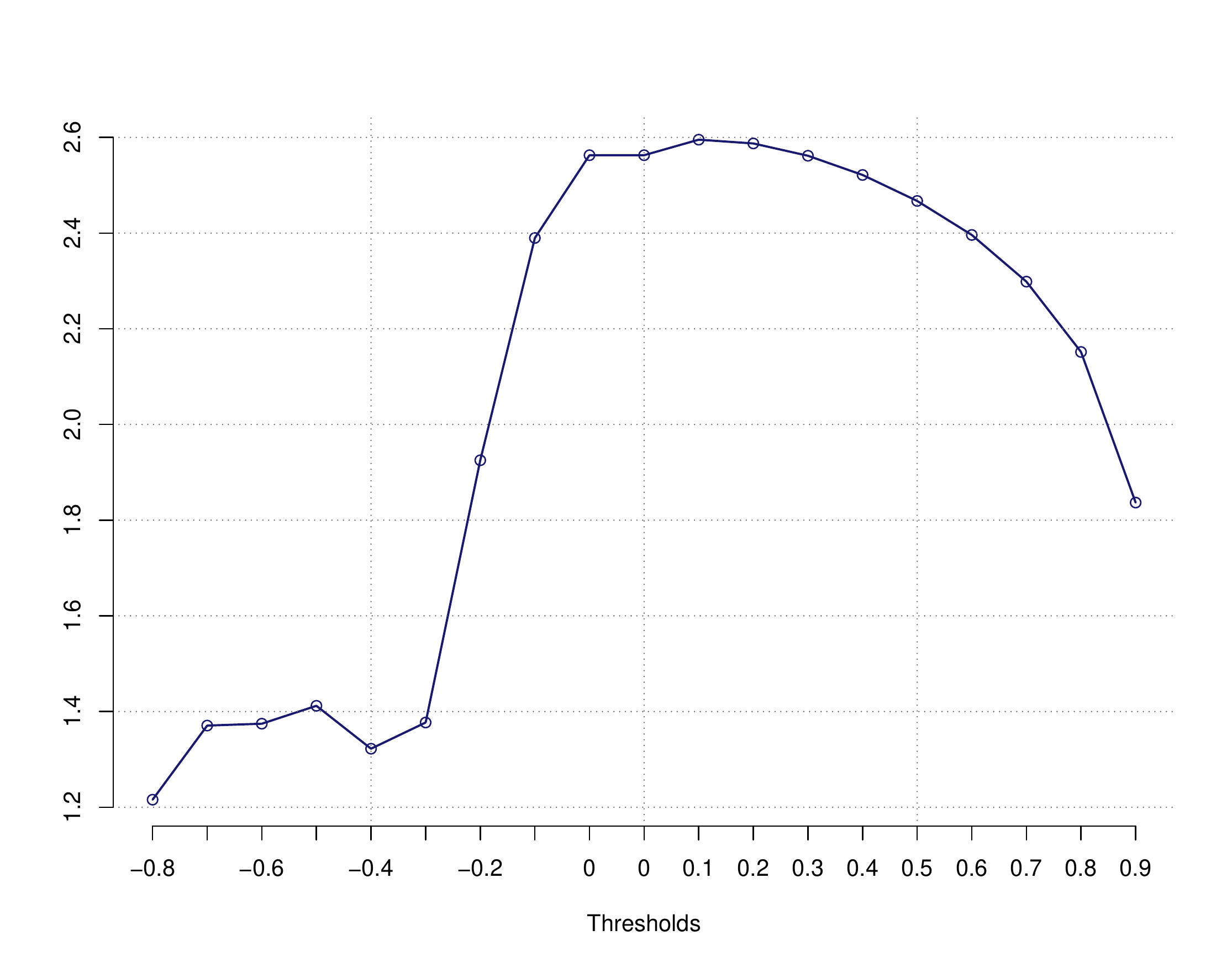}
\caption{\footnotesize Periodicity 1-year for different thresholds.}
\label{fig:11}  
\end{figure}

We performed the robust periodogram method and calculated the correlogram-based periodogram for each of the series plotted in Fig. \ref{fig:7} and Fig. \ref{fig:8}  (Fig. \ref{fig:9}   and Fig. \ref{fig:10}). All the periodograms are peaked at $1$ year (red vertical line), and this confirms the appearance of the cyclic behaviour that was found by visual inspection in Fig. \ref{fig:7} and Fig. \ref{fig:8}. In particular, the value of the robust periodogram corresponding to the annual periodicity is larger for absolute values of the threshold (Fig. \ref{fig:11}). 
Furthermore, in Fig. \ref{fig:2},  we see the annual periodicity of wind speed data - the wind velocity increases in winters and decreases in summers. Similar behaviour is observed in Fig. \ref{fig:8}  for daily connectivity density when negative correlation thresholds are considered - the density increases in winters and decreases in summers. On the other hand, for positive thresholds (Fig. \ref{fig:7}) the density increases in summers and decreases in winters.

\begin{figure}
\centering
\includegraphics[width=\linewidth]{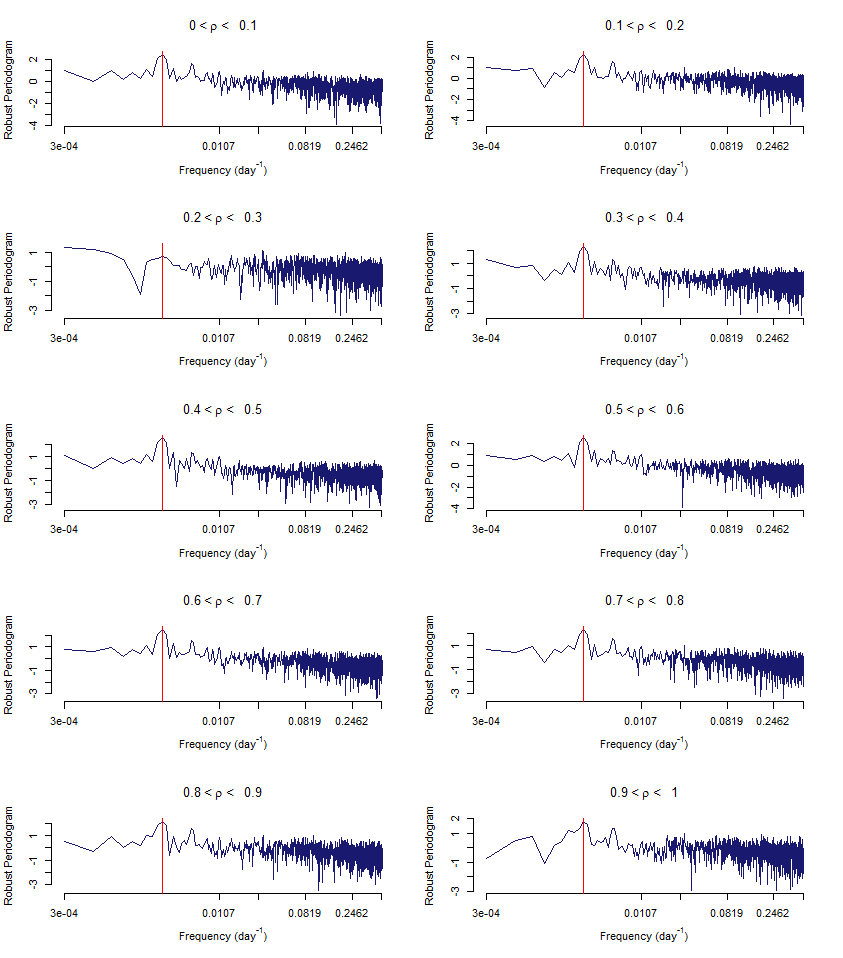}
\caption{\footnotesize Robust periodogram for different positive thresholds.}
\label{fig:13}  
\end{figure}

\begin{figure}
\centering
\includegraphics[width=\linewidth]{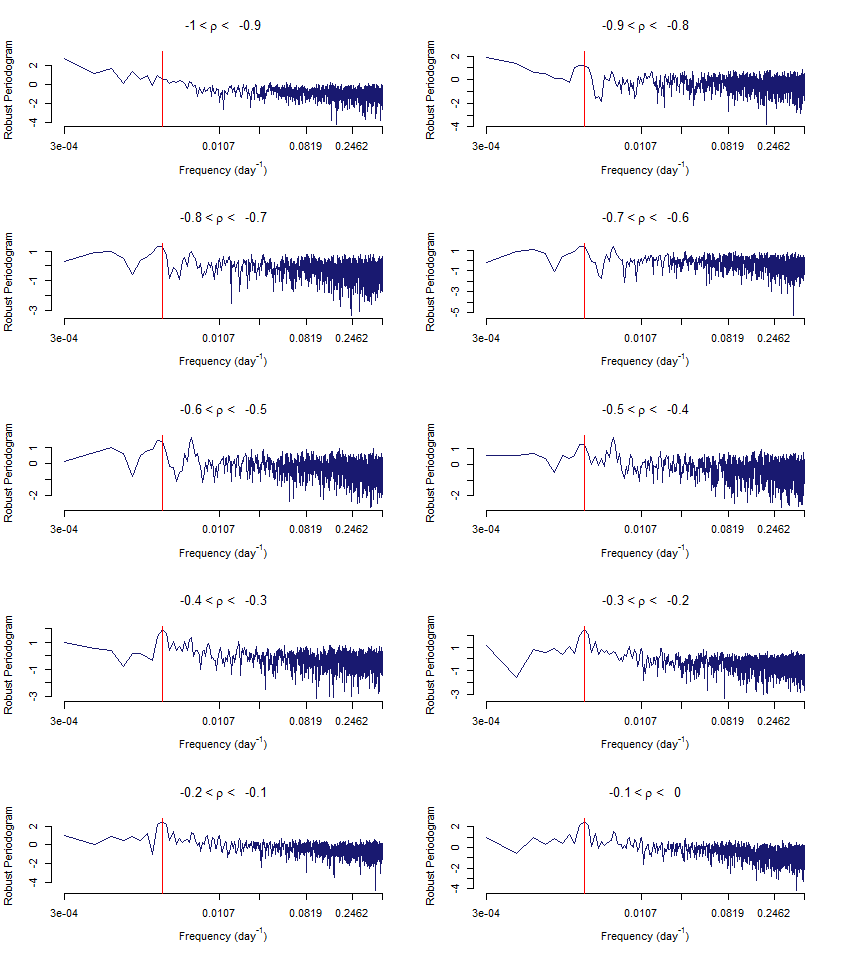}
\caption{\footnotesize Robust periodogram for different negative thresholds.}
\label{fig:14}  
\end{figure}

In order to verify if such behaviour depends on the accumulation character of the chosen threshold (the series were constructed considering all the edges whose correlation coefficient is above or below the threshold), we calculated the daily connectivity density time series considering only those edges whose correlation coefficient varies within a small range of correlation values. Fig. \ref{fig:13} and \ref{fig:14} show the robust periodograms of the daily connectivity density time series for correlation thresholds varying in the ranges $[-1,-0.9[$, $[-0.9, -0.8[$,$\ldots,$ $[0.9, 1]$. We can observe that also limiting the correlation threshold to vary within a small range, the periodograms are peaked at $1$ year.

\begin{figure}
\centering
\includegraphics[width=\linewidth]{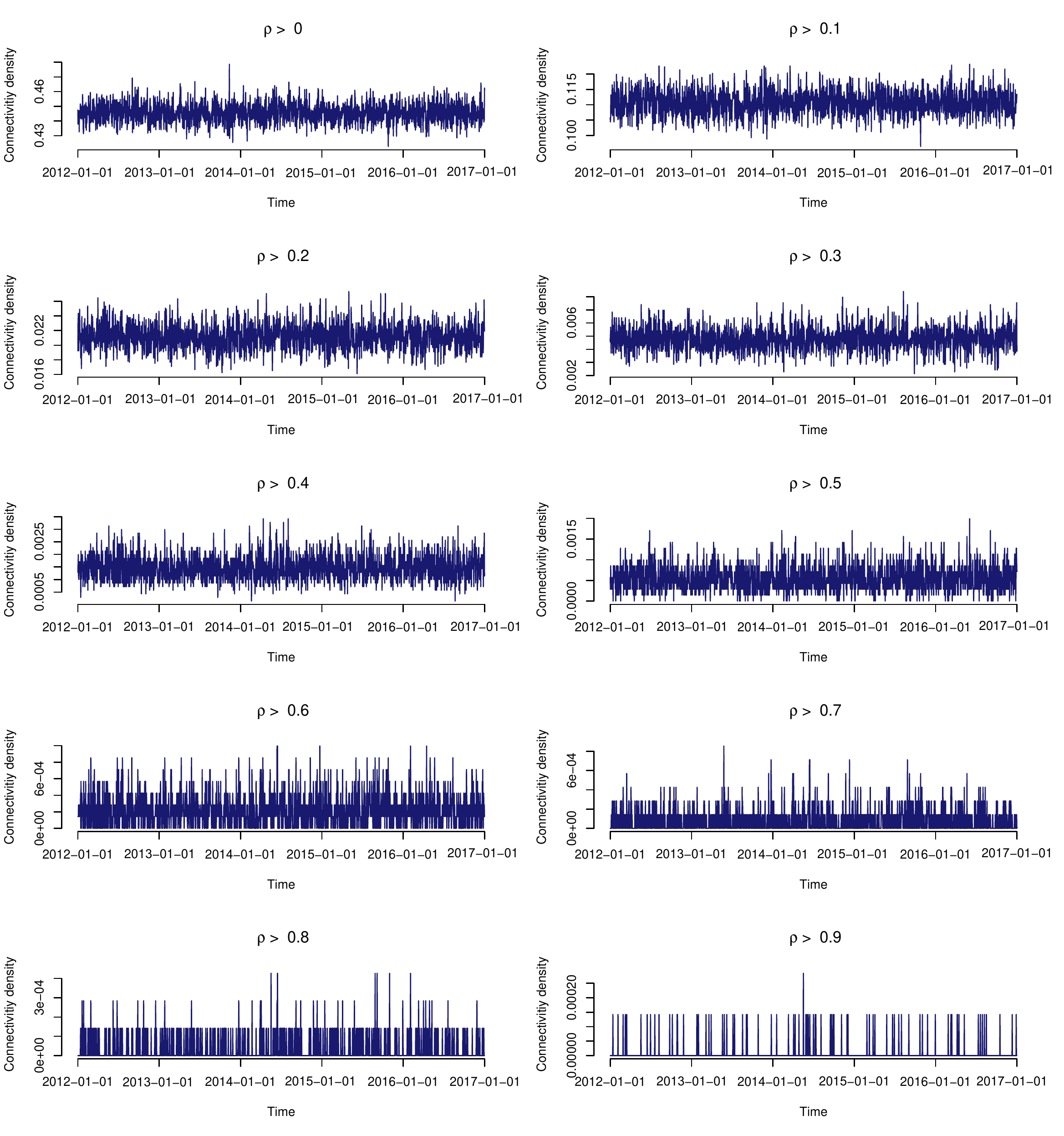}
\caption{\footnotesize Time series of daily connectivity density for simulated wind data using GEV distribution (positive correlation thresholds).}
\label{fig:19}  
\end{figure}

\begin{figure}
\centering
\includegraphics[width=\linewidth]{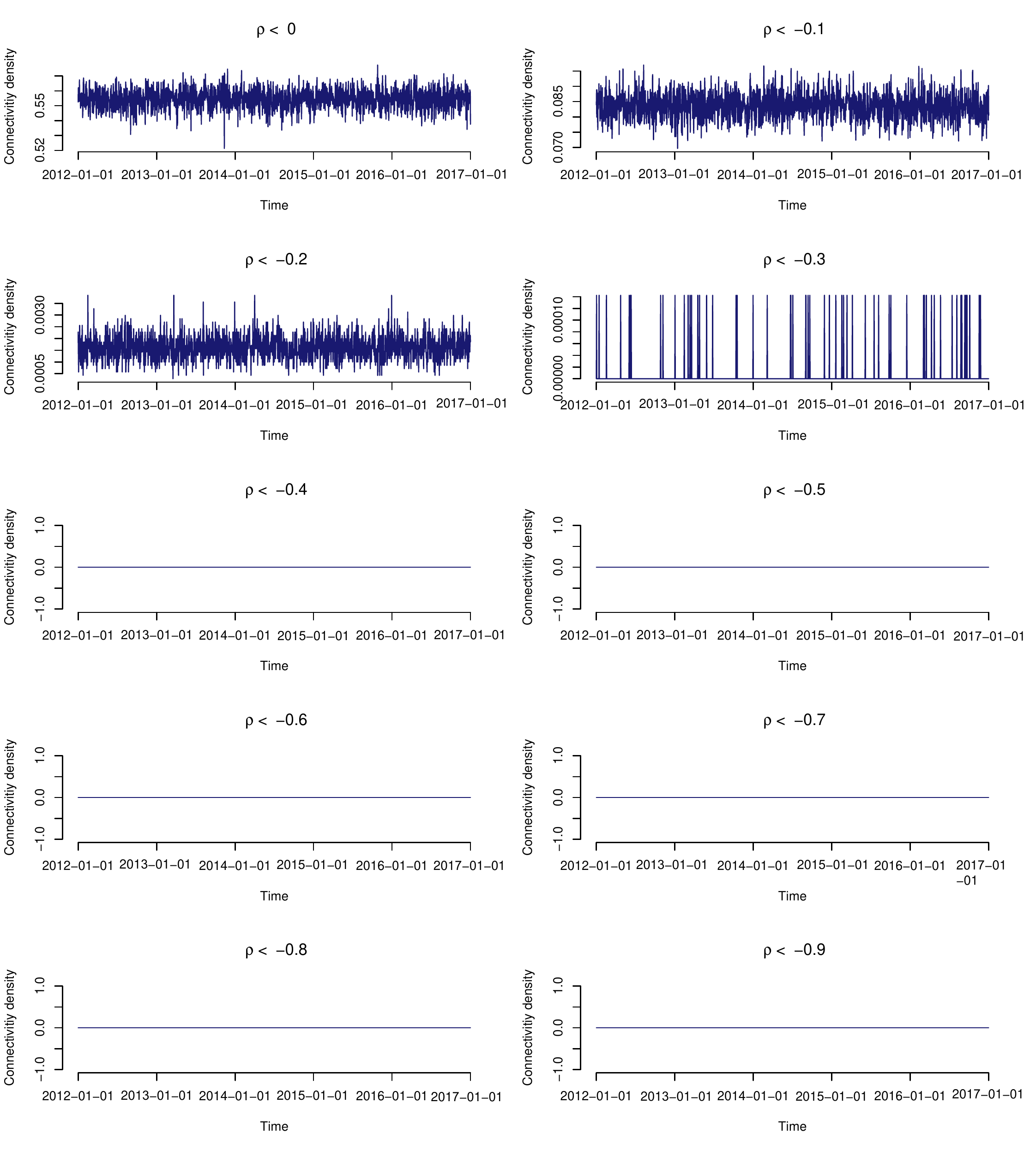}
\caption{\footnotesize Time series of daily connectivity density for simulated wind data using GEV distribution (negative correlation thresholds).}
\label{fig:20}  
\end{figure}

In order to verify if the observed periodical behaviour could be due to chance, we simulated the wind time series at each station by using the GEV distribution (since it is the best with respect to the KLD performed in section 2), although the periodic behaviour is lost, however, the range of the correlation thresholds for which the network is connected is between $-0.3$ and $0.9$ (Fig. \ref{fig:19} and Fig. \ref{fig:20}). Therefore, the GEV distribution not only models the wind speed better (strictly speaking) than the other two distributions, but it also  can reproduce the status of interactions among the wind stations, indicated by the connections among the nodes,  since the range of correlation thresholds where the interactions exist is slightly enlarged.

\section{Conclusion}
High-frequency records of $10$-min averages of wind speed at $119$ different weather-monitoring stations in Switzerland were analysed in terms of their connectivity properties. A correlation-based network was defined to explain the interactions among the stations, which were represented by nodes of a graph whose edges, connecting two nodes, were given by the Pearson correlation coefficient between the wind speed time series recorded at those nodes (stations). The selection of a threshold for the Pearson correlation coefficient leads to a selection of a number of edges whose correlation satisfies a certain relationship with the threshold (above, lower or within a given range of values), and thus, leads to a selection of a certain topology of the network, which is given by the number of nodes that are effectively interconnected. We analysed the time series of network connectivity density on a daily basis, in order to investigate the time variation of the network topology and to find possible links with environmental and/or meteo- climatic conditions in Switzerland. The main findings of our study are the following:

\begin{enumerate}

\item The daily time series of connectivity density of the wind speed network is characterized by clear annual periodicity for low absolute values of the correlation  threshold;

\item For higher absolute values of correlation threshold, not only the network is less connected (due to the lower number of edges), but the daily time series of correlation density appears more spiky, intermittent. The annual periodicity is still present, but less intensively modulates the amplitude of the connectivity density;

\item The calculation of the connectivity density for simulated wind speed time series has shown that the GEV distribution produces a connected network for thresholds between $-0.3$ and $0.9$. However, the annual periodicity is lost, which indicates that such periodicity is not generated by chance but it results from an external forcing.

\item The existence of the annual periodicity in the connectivity density seems reasonably consistent with the seasonal meteo-climatic cycle. The intensity of the annual periodicity higher for low correlation thresholds than for larger could be due to the larger cooperative effect displayed by the monitoring network at these thresholds. Lower the threshold, higher the number of nodes that are interconnected, and thus more powerful the annual periodicity that takes advantage of the cumulative effect of a higher number of interconnected nodes;

\item The lower network connectivity density at higher correlation thresholds, although does not enhance the annual periodicity, evidences the apparently intermittent character of the network’s topology, which would be probably induced by the high-frequency fluctuations of the wind speed measured at each station;

\item Two different dynamical behaviours can be recognized: one more regular with the dominance of the annual periodicity at lower correlation thresholds, and one more intermittent at higher thresholds. The regular behaviour, mainly periodic, can be considered as weather-induced, while the intermittent one can be viewed as due to the inner fluctuations of the wind speed that could be probably linked more with the local meteo-climatic conditions. In other words, the lower correlation thresholds would highlight the global weather conditions that affect more or less uniformly all the measuring stations, while those higher would enhance the local conditions that influence the climate locally, like geomorphology, sun exposure, altitude, etc.
 
\end{enumerate}
\section{Acknowledgements}
This research was partly supported by the Swiss Government Excellence Scholarships. LT thanks the support of Herbette Foundation.

The authors thank MeteoSwiss for giving access to the data via IDAWEB server. They also are grateful to the anonymous reviewers for their constructive comments that contributed to improving the paper.

\bibliography{xampl}
\bibliographystyle{elsarticle-num}


\end{document}